\documentclass[a4paper,12pt]{article}

\usepackage[T1]{fontenc}

\usepackage[utf8]{inputenc}
\usepackage{amsmath,amsfonts,epsfig,psfrag,epstopdf,array}
\usepackage{amssymb,latexsym}

\usepackage{natbib}
\setcitestyle{square} 

\usepackage{graphicx}
\graphicspath{{graphics/}}
\usepackage{fancyvrb}
\usepackage{listings}
\usepackage{bm}

\usepackage{caption}
\usepackage{subcaption}
\usepackage{chngpage}

\usepackage{setspace}

\usepackage{hyperref}

\usepackage{fancyhdr}
\usepackage{lastpage}
\usepackage[usenames,dvipsnames, table]{xcolor}
\usepackage{tabularx}

\usepackage{enumerate}

\usepackage{caption}

\usepackage{rotating}

\usepackage{caption}

\usepackage{textcomp}

\usepackage{algorithm, algorithmic}

\definecolor{gosia}{RGB}{50,200,50}
\definecolor{ejc}{RGB}{255,0,0}
\definecolor{pjs}{RGB}{0,0,255}

\newenvironment{remark}[1][Remark]{\begin{trivlist}
\item[\hskip \labelsep {\bfseries #1}]}{\end{trivlist}}

\newcommand{\qed}{\nobreak \ifvmode \relax \else
      \ifdim\lastskip<1.5em \hskip-\lastskip
      \hskip1.5em plus0em minus0.5em \fi \nobreak
      \vrule height0.75em width0.5em depth0.25em\fi}

\newcommand{\R}{\textnormal{\sffamily\bfseries R}}
\newcommand{\Rs}{\mathbb{R}}

\newcommand{\vx}{\mathbf{x} }
\newcommand{\vX}{\mathbf{X} }
\newcommand{\vT}{\mathbf{T} }
\newcommand{\vt}{\mathbf{t} }
\newcommand{\vm}{\mathbf{\mu} }
\newcommand{\vem}{\mathbf{m} }
\newcommand{\vw}{\mathbf{w} }
\newcommand{\vW}{\mathbf{W} }
\newcommand{\vS}{\mathbf{S} }
\newcommand{\vQ}{\mathbf{Q} }
\newcommand{\vP}{\mathbf{P} }
\newcommand{\vU}{\mathbf{U} }
\newcommand{\vV}{\mathbf{V} }
\newcommand{\vR}{\mathbf{R} }
\newcommand{\vL}{\mathbf{L} }
\newcommand{\N}{\mathcal{N} }
\newcommand{\vI}{\mathbf{I} }
\newcommand{\I}{\mathbf{I} }
\newcommand{\vM}{\mathbf{M} }
\newcommand{\St}{\mathcal{S}}
\newcommand{\vY}{\mathbf{Y}}
\newcommand{\vZ}{\mathbf{Z}}
\newcommand{\bmu}{\boldsymbol{\mu}}

\DeclareMathOperator*{\argmax}{arg\,max}

\newcounter{qcounter}

\newcommand*{\bigcdot}{\raisebox{-0.25ex}{\scalebox{1.5}{$\cdot$}}}

\title{Bayesian dimensionality reduction with PCA using penalized semi-integrated likelihood}

\author{Piotr Sobczyk\textsuperscript{a} ,  Ma{\l}gorzata Bogdan\textsuperscript{a,b} , Julie Josse\textsuperscript{c,d}}
\date{}

\begin{document}

\maketitle

{\centering
\vspace*{-0.3cm}
\noindent\textsuperscript{a} Faculty of Pure and Applied Mathematics,
Wroc{\l}aw University of Science and Technology, Wroclaw,
Poland\\
\textsuperscript{b} Institute of Mathematics, University of Wroc{\l}aw, Wroclaw, Poland\\
\textsuperscript{c} INRIA, Paris, France \\
\textsuperscript{d} Department of Applied Mathematics, Agrocampus Ouest, Rennes, France\par\bigskip }

\begin{abstract}

We discuss the problem of estimating the number of principal components in Principal Components Analysis (PCA). Despite of the importance of the problem and the multitude of solutions proposed in the literature, it comes as a surprise that there  does not exist a  coherent asymptotic framework which would  justify different approaches depending on the actual size of the data set. In this paper we address this issue by presenting an approximate Bayesian approach based on Laplace approximation and introducing a general method for building the model selection criteria, called PEnalized SEmi-integrated Likelihood (PESEL). Our general framework  encompasses a variety of existing approaches based on probabilistic models, like e.g.  Bayesian Information Criterion for the Probabilistic PCA (PPCA), and allows for construction of new criteria, depending on the size of the data set at hand and additional prior information. Specifically, we apply PESEL to derive two new criteria for data sets where the number of variables substantially exceeds the number of observations, which is out of the scope of currently existing approaches.
We also report results of extensive simulation studies and real data analysis, which illustrate good properties of our proposed criteria as
compared to the state-of-the-art methods and very recent proposals. 
Specifically, these simulations show that PESEL based criteria can be quite robust against  deviations from the probabilistic model assumptions.
Selected PESEL based criteria for the estimation of the number of principal components are implemented in R package \emph{varclust}, which is available on github
(\url{https://github.com/psobczyk/varclust}).


\end{abstract}

\noindent%
\vfill

\newpage
\doublespacing

\section{Introduction}

\noindent Principal Component Analysis (PCA)~\citep{pearson} is a widely used technique for dimensionality reduction. 
It is applied in many fields as a way to visualize multidimensional data, which is projected onto a number of orthogonal
directions in a low dimensional space. Researcher can examine this representation to get intuition about the data structure 
and conjecture a hypothesis, that would be 
further investigated. We treat first components in PCA as signal, and the rest of components as noise. 
In such exploratory analysis, it is often redundant to have a tool for precise choice of the number of non-noise components. 
But there are many situations where accuracy of this estimation is important. 
For example in projective clustering (see e.g.~\citep{ksubspaces}), where data is clustered along
various linear subspaces, 
an incorrect estimation of subspaces dimensions may lead to the choice of wrong number of clusters and incorrect segmentation. 
Other example is the important problem of missing values in PCA,
where inaccurate estimation of the number of components may lead to overfitting
(see~\citep{josse2009gestion},~\citep{Ilin_practicalapproaches},~\citep{Josse2011}).

\noindent Let $\vX=[x_{ij}]_{n\times p}$ be high-dimensional data, where $n$ is the number of observations and
$p$ is the number of variables. Consider the fixed effect model for PCA:
\begin{equation}\label{pca_general_model_hoff}
 \vX - \bmu_{n \times p} = \vM + E,
\end{equation}
where $\bmu$ is of rank one, $\vM$ is assumed to be of low rank $k\leq \min(n,p)$ and $E=[\epsilon_{i,j}]_{n \times p}$ is a matrix of i.i.d. errors,
$\epsilon_{i,j} \sim \N(0, \sigma^2)$. 
Equivalently we may write model ~(\ref{pca_general_model_hoff}) as in ~\cite{fixedPCA}:
\begin{align}\label{general_model_pca}
 \vX - \boldsymbol{\mu}_{n \times p} = \vT \vW^T + E,
\end{align}
where  
$\vT = [t_{i,l}]_{n \times k}$ is a matrix whose columns contain factors spanning the data, and 
$\vW = [w_{i,l}]_{p \times k}$ is a matrix of coefficients.


\noindent Given the number of components $k$, maximum likelihood estimators for $\vT$ and $\vW$ are 
obtained by performing SVD decomposition of $\vX - \boldsymbol{\mu}$ truncated at the order $k$~(see for example~\citep{fixedPCA}, ~\citep{gls}).

\noindent In~\cite{jolliffe2002principal} three types of methods for choosing the number of factors  are distinguished.
First are ad-hoc rules such as a scree test~\citep{screeTest} or a rule of thumb  that chooses the smallest number of 
factors which jointly explain e.g. 90\% of variance of data.
Although these methods are usually fast and easy to implement, they are difficult to use in automatic way, since in high-dimensional data it is common that few first components explain a lot of variance even if data
is entirely  random~\citep{exploratory:multivariate:R}. Therefore these heuristic approaches are rather useless when PCA 
is used without supervision.

\noindent Second type are techniques that view the problem in a more systematic way but are not based on any probabilistic 
assumptions. Among them are bootstrap and permutation methods (see~\citep{jackson1993stopping}) and cross-validation (see~\citep{owen2009}, ~\citep{Josse20121869}).
Finally there exists a group of methods based on asymptotic approximation for a specific probabilistic model, which we discuss in more detail in Section~\ref{sec:probabilistic_approaches_overview}.

\noindent Despite the fact that there exist various probabilistic approaches 
for choosing the number of principal components~(e.g.~\citep{Tipping99probabilisticprincipal}),
there seems to be no coherent framework that would unify them. 
Additionally, according to our knowledge, there exist no proper criteria to deal with the situation when the number of variables is substantially larger than the number of observations. 
In this paper we address these issues 
by introducing a novel approximate Bayesian framework for model selection.
As existing full Bayesian methods (see e.g.~\citep{Hoff07modelaveraging}) are computationally expensive and require specifying many prior distributions, 
we suggest instead to use approximated formula for the posterior probability. 
Our focus is on Bayesian Information Criterion, which is based on the Laplace approximation.
This approach requires the number of degrees of freedom in the model to be independent of the number of observations. As this is not satisfied by
(\ref{general_model_pca}), we need to analytically integrate out some of the priors. First strategy relies on integrating out the prior on $\vT$, 
which reduces number of parameters in the model, so it no longer depends on $n$. 
Hence we get PESEL (PEnalized SEmi-integrated Likelihood) valid for $n\to \infty$. 
Second strategy is integrating out the prior on $\vW$, which gives criterion valid for $p\to \infty$. 

\noindent Thus, it is  useful for the sequel to express~(\ref{general_model_pca}) as a model for either rows or columns of the matrix $\vX$,
for which we shall use notation $x_{i \bigcdot}$ and $x_{\bigcdot j}$ respectively.

\begin{align}\label{fixed_rows_columns_pca}
\begin{split}
\vx_{i \bigcdot} - \boldsymbol{\mu}_{i \bigcdot} &= \sum_{l=1}^k t_{i,l} \, \vw_{\bigcdot  l} + \epsilon_{i \bigcdot} = \vt_{i \bigcdot} \, \vW^T + \epsilon_{i \bigcdot},
\; \; \; i = 1, \dots, n, \;\; \; \epsilon_{1 \bigcdot}, \dots ,\epsilon_{n \bigcdot} \text{ are i.i.d. } \N(0, \I_p), \\
\vx_{\bigcdot  j} - \boldsymbol{\mu}_{\bigcdot  j}&= \sum_{l=1}^k w_{j,l} \, \vt_{\bigcdot  l}+ \epsilon_{\bigcdot j} = \vT \, w^T_{j \bigcdot} + \epsilon_{\bigcdot j} ,
\; \; \; j = 1, \dots, p , \;\; \; \epsilon_{\bigcdot 1}, \dots ,\epsilon_{\bigcdot p} \text{ are i.i.d. } \N(0, \I_n).
\end{split}
\end{align}

\noindent Later in Section~\ref{sec:probabilistic_approaches_overview}, we shall assume 
 for the model for rows $\vx_{i \bigcdot}$, that 
$\boldsymbol{\mu}= \begin{bmatrix}
    \mu_{1} & \mu_{2} & \dots  & \mu_{p} \\
    \vdots & \vdots & \vdots  & \vdots \\
    \mu_{1} & \mu_{2} & \dots  & \mu_{p} \\
\end{bmatrix}$.

\noindent Analogously, for the model for columns $\vx_{\bigcdot j}$, we shall assume that
$\boldsymbol{\mu} = \begin{bmatrix}
    \mu_{1} & \dots  & \mu_{1} \\
        \mu_{2} & \dots  & \mu_{2} \\
    \vdots & \vdots & \vdots  \\
    \mu_{p} & \dots  & \mu_{p} \\
\end{bmatrix} $. In this case vector $[ \mu_{1}, \dots, \mu_{p}]$ can be interpreted as the mean of the distribution generating variables rather than observations.
Such a model is not common, yet it naturally occurs when $p$ is much larger than $n$. In this case  all $p$ variables are spanned by just few factors and each new variable is simply defined by  the vector of corresponding coefficients, which can be modeled as randomly selected from some underlying distribution.

\noindent The outline of the paper is as follows. In Section~\ref{sec:probabilistic_approaches_overview}, we  introduce the estimation of the number
of principal components in a Bayesian setting.
In Section~\ref{sec:ppca_bic} we formulate PESEL for the prior on $\vT$ and in Section~\ref{sec:big_p} 
we do the same for the prior on $\vW$. 
In Section~\ref{sec:simulation}, we assess the performance of the respective selection criteria
by computer simulations, in which we compare PESEL to existing and very recents methods.
We focus on robustness against deviations from the assumed probabilistic model.
In Section~\ref{sec:real_data}, we present the analysis of the real data, which additionally confirms 
soundness of the PESEL methodology.
Our criteria are implemented in the R package \emph{varclust} and the code to reproduce all the results available in this paper (including implementation of existing methods for which codes were not provided by the authors)  is available at \url{https://github.com/psobczyk/pesel_simulations}.

\section{Methods based on probabilistic model}\label{sec:probabilistic_approaches_overview}

\noindent To estimate $k$ in a probabilistic setting, we focus on using
Bayesian approach and view it as a problem of model selection, 
where each model $M$ specifies the number of principal components $k$.
In general, following maximum a posteriori (MAP) rule, we want to choose a model that is most probable 
 given the data. 
 To do that, we maximize logarithm of the posterior probability
 \begin{align*}
  \log(P(M|\vX)) &= \log(P(\vX|M)) + \log(P(M)) + C(\vX) \nonumber \\
  &= \log\left(\int_{\Theta} p(\vX | \theta) \, \pi_M(\theta) \, d\theta \right) + \log(P(M)) + C(\vX),
 \end{align*}
  where $P(M)$ is a prior distribution on considered models,
  $\pi_M(\theta)$ is a prior distribution on parameters in a model $M$, and 
  $C(\vX)$ is a scaling factor that does not depend on $M$.
  In the reminder of this paper we assume that $P(M)$ is uniform i.e. it does not influence model selection.
  However; the method can be easily extended for any informative prior distribution on $M$.
  
  \noindent In terms of model~(\ref{general_model_pca}), $\theta=(\bmu, \vW, \vT, \sigma) \in \Theta$,
  and $P(\vX|M)$ takes the following form:
  
  \begin{align}\label{posterior_as_integral}
   \log(P(\vX|M)) = \log \int_{\Theta} p(\vX|\bmu, \vW, \vT, \sigma) \, \pi_M(\bmu, \vW, \vT, \sigma) \, d\bmu \, d\vW \,d\vT \, d\sigma.
  \end{align}
  
\subsection{Full Bayesian approach}

\noindent  There exist several Bayesian methods for estimating the number of principal components in
the PCA model.
One of them was proposed in~\cite{Bishop:1999:BP:340534.340674}, who used
the following priors in model~(\ref{general_model_pca}):

\begin{minipage}{.2\linewidth}
\begin{flushright}
 
\end{flushright}
\end{minipage}
\begin{minipage}{.25\linewidth}
\begin{flushright}
 \begin{align}
 \vt_{i \bigcdot} &\sim \N(0, \vI), \nonumber \\
 \vw_{j \bigcdot} &\sim \N(0, \frac{1}{\alpha_j} \vI), \nonumber
\end{align} 
\end{flushright}
\end{minipage}
\begin{minipage}{.35\linewidth}
\begin{flushleft}
  \begin{align}\label{eq:priors_bishop}
   \alpha_j &\sim \Gamma(a_{\alpha}, b_{\alpha}),  \\
 \frac{1}{\sigma^2} &\sim \Gamma(c_{\sigma}, d_{\sigma}), \nonumber
\end{align}
\end{flushleft}
\end{minipage}

\noindent where $a_{\alpha}, b_{\alpha}, c_{\sigma}, d_{\sigma}$ are model hyperparameters.
Rows of $\bmu$ were estimated with $\bar{\vx} = (\overline{\vx}_{\cdot 1}, \dots, \overline{\vx}_{\cdot p})$, 
where $\overline{\vx}_{\cdot j} = \frac1n \sum^n_{i=1} x_{ij}$.

\noindent  \citet{Bishop:1999:BP:340534.340674} introduces non-discrete ``model selection'' for PCA, 
by the means of continuous parameters, that control variability of columns of $\vW$. More specifically, large value
of $\alpha_j$ effectively ``switches off'' $\vw_{j \bigcdot}$.
\citet{Bishop:1999:BP:340534.340674} proposes three computational methods for marginalizing over the posterior
on $\vW$, including among others Markov Chain Monte Carlo.
In the follow up paper, \citet{bishop1999variational} recommends the variational approach, which
proves to be the most efficient.
This idea was further pursued by~\citet{Ilin_practicalapproaches}, 
who propose fast algorithm for variational Bayesian PCA (VBPCA), which
is an extension of regular EM algorithm for maximizing likelihood function~\citep{Dempster77maximumlikelihood}.
However, as mentioned before, VBPCA does not enable direct estimation of the number of PCs. 

\noindent Another full Bayesian approach was proposed by~\citet{Hoff07modelaveraging}, who
considered the model~(\ref{pca_general_model_hoff}) with $\bmu=0$ and
priors imposed on components of SVD decomposition of matrix $\vM = \vP \vL \vQ^T$:

\begin{minipage}{.1\linewidth}
\begin{flushright}
 
\end{flushright}
\end{minipage}
\begin{minipage}{.35\linewidth}
\begin{flushright} 
 \begin{align} 
\vP &\sim uniform(\St_{k\times n}), \nonumber \\
\vQ &\sim uniform(\St_{k\times p}), \nonumber \\ 
 \epsilon_{i \bigcdot}  &\sim \N(0, 1/\phi), \nonumber \\ 
 l_{i,i} &\sim \N(\ell, 1/\psi), \nonumber 
\end{align} 

\end{flushright}
\end{minipage}
\begin{minipage}{.35\linewidth}
\begin{flushleft}
  \begin{align} \label{hoff_probabilistic_model}
\ell&\sim \N(l_0,v_0^2),\\
\psi&\sim \Gamma(\eta_0/2, \eta_0\tau_0^2/2),\nonumber  \\
\phi&\sim \Gamma(\nu_0/2, \nu_0\sigma_0^2/2), \nonumber
\end{align}
\end{flushleft}
\end{minipage}

\noindent where $uniform(\St_{k\times n})$ denotes the uniform distribution on the 
Stiefel manifold of orthogonal matrices~\citep{chikuse2003statistics},
$l_{i,i}$ are elements of diagonal matrix $L$ and $(l_0,v_0^2)$, $(\eta_0, \tau_0)$, $(\nu_0,\sigma_0)$ are model hyperparameters.
To estimate the number of principal components, \citet{Hoff07modelaveraging}  considers the model with $k=p$ and uses the prior on $l_{i,i}$ specified in (\ref{hoff_probabilistic_model}) as a continuous component in the spike and slab prior, with a positive mass at 0.
Posterior distributions for parameters are computed with MCMC.
Software provided by~\citet{Hoff07modelaveraging} requires $n\geq p$; however, because of symmetry in the
model~(\ref{hoff_probabilistic_model}), one may transpose the data and then use the method when $p>n$.
Due to the complexity of MCMC, implementation is rather slow and does not scale very well. 
Because even for moderately sized matrices (i.e. $1000\times 100$) generating Markov chain of the 
length 1000  takes more than two hours,
we decided not to include this method in the simulation study.

\subsection{PESEL -- PEnalized SEmi-integrated Likelihood} \label{sec:pesel}

Exact calculation of posterior probability, as in~\cite{Hoff07modelaveraging}, is computationally expensive
and requires specifying all prior distributions.
Instead one can perform model selection by approximating integral~(\ref{posterior_as_integral}) using 
Laplace approximation (for more details about Laplace approximation see Appendix~\ref{app:laplace}). 
The integrated likelihood in the fixed effect models~(\ref{fixed_rows_columns_pca}) for rows takes the form:

\begin{align}\label{posterior_likelihood_pca}
   \log{P(\vX|M)} &= \log \int_{\Theta} P(\vX|\bmu, \vW, \vT, \sigma) \, 
   \pi_M(\bmu, \vT, \vW, \sigma) \, d\bmu \, d\vT \, d\vW \, d\sigma \nonumber \\
   &= \log \int_{\Theta} \Pi_{i=1}^{n} \phi(x_{i \bigcdot}; \bmu_{i \bigcdot} + \vt_{i \cdot} \vW^T, \sigma^2 \vI_p) \, 
   \pi_M(\bmu, \vT, \vW, \sigma) \, d\bmu \, d\vT \, d\vW \, d\sigma,
\end{align}
where $\phi(x; \vem, \boldsymbol{\Sigma})$ is a probability density function of normal distribution with mean $\vem$ and covariance matrix $\boldsymbol{\Sigma}$.
It is invalid to apply Laplace approximation directly to the integral~(\ref{posterior_likelihood_pca}) as
number of parameters in this model is proportional to both the number of observations $n$ and the number of variables $p$.
This violates Laplace approximation assumption that dimension of the parameter space is constant.
Thus, to perform approximation one should reduce dimensionality, for example by integrating out the 
prior on either $\vT$ or $\vW$.
This choice is determined by asymptotics. For $p \to \infty$  we need to integrate out $\vW$ because
its number of parameters grows linearly in $p$. 
Similarly, for $n \to \infty$ $\vT$ needs to be integrated out. After integrating out one of the priors,
we can apply the Laplace approximation for the resulting semi-integrated likelihood. This yields
a new Bayesian criterion for estimating the model dimension, which we call PEnalized SEmi-integrated
Likelihood (PESEL).

\subsubsection{PESEL for $p$ fixed and $n \to \infty$} \label{sec:ppca_bic}

\noindent If we work in asymptotic regime when $n \to \infty$, then to apply Laplace approximation we need to integrate out $\vT$ from~(\ref{posterior_likelihood_pca}) according to the formula:

  \begin{align}
   \log{P(\vX|M)} &= \log \int SIL(\vX|\mu, \vW, \vT, \sigma) \pi(\mu, \vW, \sigma) \, d\mu \, d\vW \, d\sigma, \nonumber
   \end{align}
where $SIL(\vX|\mu, \vW, \vT, \sigma):= \int P(\vX|\mu, \vW, \vT, \sigma) \, \pi(\vT) \, d\vT$ is a semi-integrated likelihood
function. In the above $\mu =  \bmu_{i \bigcdot} = [\mu_{1}, \mu_{2}, \dots, \mu_{p} ]$ are rows of $\bmu$ from equation~(\ref{general_model_pca}).

We propose using two forms of PESEL, based on specific priors on rows of scores matrix $\vT$.
Firstly, we use the prior $\vt_{i, \bigcdot} \sim \N(0, \I_k)$, which brings the  
Probabilistic Principal Component Analysis (PPCA)  model of~\citet{Tipping99probabilisticprincipal} 
(a random-effects version of our fixed-effects model \eqref{general_model_pca}). 
In this case $\vt_{i \bigcdot} \, \vW^T \sim \N(0, \vW \vW^T)$. Therefore our 
semi-integrated likelihood is reduced to the likelihood in PPCA, under which $ \vx_{1 \bigcdot},\ldots,  \vx_{n \bigcdot}$ are independent and
\begin{align} \label{normal_prior_t}
   \vx_{i \bigcdot} = \mu + \vt_{i \bigcdot} \vW^T + \epsilon_{i \bigcdot} \sim \N(\mu; \vW \vW^T + \sigma^2 \vI_p).
\end{align}


\noindent Second approach is to consider prior $\vt_{i \bigcdot} \sim \frac{1}{\beta} \N(0, \I_k)$ with
the additional restriction $\vW^T \vW = \vI_k$.
This constraint makes all singular values in PCA \textit{homogeneous} (we will refer to it with the notation \textit{homo}). 
In other words, all PCA factors are equally weighted, i.e. none of directions dominates the data.
This distinguishes it from the previous prior, which allows \textit{heterogeneous} (noted \textit{hetero}) singular values. 
Such a {\it homogeneous} distribution for $\vt_{i \bigcdot}$ was discussed in~\citep{637226}.
With this prior $\vt_{i \bigcdot} \, \vW^T \sim \N(0, \frac{1}{\beta} \vW \vW^T)$ and the 
semi-integrated likelihood function for rows of $\vX$ corresponds to $ \vx_{1 \bigcdot},\ldots,  \vx_{n \bigcdot}$ being independent and
\begin{equation}\label{normal_variance_prior_t} 
  \vx_{i \bigcdot} \sim \N(\mu; \frac{1}{\beta} \vW \vW^T + \sigma^2 \vI_p).
\end{equation}

\noindent Let us now focus on the semi-integrated likelihood specified in formula~(\ref{normal_prior_t}), which yields 

  \begin{align}\label{posterior_likelihood_ppca}
   \log{P(\vX|M)}
   &= \log \int  SIL(\vX|\mu, \vW, \sigma) \, \pi(\mu, \vW, \sigma) \, \; d\mu \, d\vW\, d\sigma  \nonumber \\
   &=  \log \int \Pi_{i=1}^{n} \phi(x_{i \cdot} - \mu; \vW \vW^T + \sigma^2 \vI_p) \, \pi( \mu, \vW, \sigma) \, d\mu \, d\vW\, d\sigma .
  \end{align}

\noindent Now, assuming that $p<<n$ and provided that $\pi(\mu, \vW, \sigma)$ satisfies standard regularity conditions, 
it is possible to apply Laplace approximation to the integral~(\ref{posterior_likelihood_ppca}):

\begin{equation}\label{model_M_laplace_formula}
 \log{P(\vX|M)} \approx \log{SIL(\vX| \hat{\mu}, \hat{\vW}, \hat{\sigma})} - \frac{1}{2} K \log n,
\end{equation}
where $K=\cfrac{pk-\frac{k (k+1)}{2} + k + p + 1}{2}$ is the 
number of free parameters in the integral~(\ref{posterior_likelihood_ppca})
(details can be found in Appendix~\ref{app:number-parameters}).

\noindent From~\cite{Tipping99probabilisticprincipal}, we get parameter values that 
maximize semi-integrated likelihood $SIL$:
\begin{align} \label{ml_estimates_PPCA}
\hat{\mu} &= \frac1n \sum_{i=1}^{n} \vx_{i \bigcdot}, \\
 \hat{\vW} &= \vU (\boldsymbol{\Lambda_k} - \sigma^2 \I_k)^{1/2} \vR, \\
  \hat{\sigma}^2 &= \frac{\sum_{j=k+1}^p \lambda_j}{p-k}\;\;, \nonumber
\end{align}
where orthogonal matrix $\vU$ contains top $k$ eigenvectors of the sample covariance matrix $\boldsymbol{\Sigma} = \frac{\vS}{n}$ with
$\vS = \sum_{i=1}^{n} \left(\vx_{i \bigcdot}  -\bar{\vx} \right)^T (\vx_{i \bigcdot} -\bar{\vx})$,
$\boldsymbol{\Lambda_k}= \left( \begin{array}{ccc}
\lambda_1 & \dots & 0 \\
\vdots & \lambda_l & \vdots \\
0 & \dots & \lambda_k \end{array} \right) $ contains corresponding eigenvalues, and
$\vR$ is a rotation matrix.

\noindent Plugging in ML estimates~(\ref{ml_estimates_PPCA}) to semi-integrated likelihood, after some algebra, 
we get~(see also~\citep{Minka00automaticchoice}):
  \begin{equation}\label{maximized_likelihood}
SIL(\vX| \hat{\mu}, \hat{\vW}, \hat{\sigma}^2) = 
    (2\pi)^{-pn/2} \left( \Pi_{j=1}^{k} \lambda_j  \right)^{-n/2} (\hat{\sigma}^2)^{-n(p-k)/2} \exp ( -\frac{pn}{2}), 
  \end{equation}

\noindent which together with (\ref{model_M_laplace_formula}) gives  Penalized SEmi-integrated Likelihood criterion (PESEL):
\begin{equation}\label{BIC_model_PPCA}
  PESEL^{hetero}_n
  = -\frac{pn}{2} \log{2\pi} -\frac{n}{2} \sum_{j=1}^{k} \log \lambda_j  - 
  \frac{n (p-k)}{2} \log(\hat{\sigma}^2) -\frac{pn}{2} - \log(n) \frac{pk-\frac{k (k+1)}{2} + k + p + 1}{2}.
\end{equation}

\begin{remark}\ \\
\noindent $PESEL^{hetero}_n$ coincides with BIC for PPCA, as proposed by~\citet{Minka00automaticchoice}. 
The major difference is that~\citet{Minka00automaticchoice} developed this criterion using specific prior distribution on $\vW$
and noise $\sigma^2$,
while we show that the approximation is valid for any regular prior on these parameters.
\citet{Minka00automaticchoice} also suggests  a second criterion called Laplace evidence, 
which depends on selected prior distribution on $W$.
This idea was further extended by~\citet{hoyle2008automatic}, who added additional terms in the approximation,
which allows to deal with the situation when $p$ increases proportionally to $n$. However the drawback of this approach is that 
it is highly dependent on the prior on $W$ and does not solve the problem when $n=const$ and $p \to \infty$, which is the main 
focus of this article and is solved by the $PESEL_p$ introduced in the next section.
\end{remark}

\noindent Consider now semi-integrated likelihood in~(\ref{normal_variance_prior_t}). 
As before, we can compute parameters that maximize semi-integrated likelihood:
\begin{align} \label{ml_estimates_Rayner}
\hat{\mu} &= \frac1n \sum_{i=1}^{n} \vx_{i \bigcdot}, \nonumber \\
 \hat{\vW} &= \text{top $k$ eigenvectors of covariance matrix},  \\
 \hat{\sigma}^2 &= \frac{\sum_{j=k+1}^n \lambda_j}{n-k}, \nonumber \\
 \hat{\beta} &= \frac{\sum_{j=1}^k \lambda_j}{k} - \hat{\sigma}^2. \nonumber \\
\end{align}

\noindent Then PESEL is of the form:
\begin{align}\label{BIC_model_Rayner}
  PESEL^{homo}_n = 
  -\frac{pn}{2} \log{2\pi} -\frac{n k}{2} \log\left(\frac{\sum_{j=1}^k \lambda_j}{k}\right) - 
  \frac{n (p-k)}{2} \log(\hat{\sigma}^2) \\ -\frac{pn}{2} - \log(n) \frac{pk-\frac{k (k+1)}{2} + p + 1 + 1}{2}.\nonumber
\end{align}

\begin{remark}\ \\
\noindent As for $PESEL^{homo}_n$, it uses the prior and marginal likelihood from~\citep{637226}. 
However, \citet{637226} did not penalize likelihood for the number of parameters. 
Thus their criterion tends to significantly overestimate the number of components, which was confirmed in simulations.

\noindent Let us give some insight on the difference between two priors and criteria presented in this Section.
Observe that in~(\ref{BIC_model_PPCA}) there is a term with a sum of logarithms of first $k$ eigenvalues $\sum_{j=1}^{k} \log \lambda_j$. 
As in model~(\ref{normal_variance_prior_t}) $\vW$ is assumed to be
orthonormal, all $k$ largest eigenvalues have to be equal, and their estimate is
$\frac{\sum_{j=1}^{k} \lambda_j}{k}$.
Thus, in the corresponding term in~(\ref{BIC_model_Rayner}), sum of logarithms of $k$ largest eigenvalues is 
$k \log\left(\frac{\sum_{j=1}^k \lambda_j}{k}\right)$. 
That observation is yet another justification for referring to formula~(\ref{BIC_model_PPCA}) as \textit{heterogeneous} PESEL and
to formula~(\ref{BIC_model_Rayner}) as \textit{homogeneous} PESEL.
\noindent The other difference is in the penalty term. Because of the eigenvalue equality assumption in \textit{homogeneous} PESEL the number of free parameters related to the eigenvalue estimation is equal to 1, while in {\it heterogeneous} PESEL this number is equal to $k$, for $k$ distinct eigenvalues we need to estimate. 
\end{remark}

\subsubsection{PESEL for $n$ fixed and $p \to \infty$} \label{sec:big_p}

The case of asymptotics  with $p \to \infty$ and $n=const$, which is of great interest in many applications, was as far as we know, never properly discussed.
In this setting we need to integrate out $\vW$ from~(\ref{posterior_likelihood_pca}).
Then it becomes possible to apply Laplace approximation.
Consider the fixed-effects model (\ref{fixed_rows_columns_pca}), expressed in terms of columns of matrix $\vX$:

\begin{equation}
 \vx_{\bigcdot  j}  \sim \N( \vm + \vT \vw^T_{j \bigcdot},  \sigma^2 \vI_n), \nonumber
\end{equation}
where $\mu =  \bmu_{\bigcdot  j} = [\mu_{1}, \mu_{2}, \dots, \mu_{n} ]^T$ is the column of $\bmu$ from equation~(\ref{general_model_pca}).

Analogously to previous section, we propose using one of two priors on rows of loadings matrix $\vW$.
The difference between priors was previously described in Section~\ref{sec:ppca_bic}.

$\vw_{j \cdot} \sim \N(0, \I_k)$, which yields: 
\begin{equation} \label{normal_prior_w}
   \vx_{\bigcdot  j} \sim \N(\vm; \vT \vT^T + \sigma^2 \vI_n).
\end{equation}

$\vw_{j \cdot} \sim \frac{1}{\beta} \N(0, \I_k)$ with constraint that $\vT^T \vT = \vI_k$, which yields:
\begin{equation}\label{normal_variance_prior_w} 
  \vx_{\bigcdot  j} \sim \N(\vm; \frac{1}{\beta} \vT \vT^T + \sigma^2 \vI_n).
\end{equation} 

\noindent For both priors marginal distributions for variables $\vx_{\bigcdot  j}$ are independent with
covariance matrix depending only on factors $\vT$.
The related mixture model with random loadings $\vW$ and fixed factors $\vT$ is in fact interpretable and intuitive. This is because when $p$ is much larger than $n$, we may model our variables as randomly selected from the set of linear combinations of the small number $k\leq n$ of fixed factors.
Now, observe that probabilistic models (\ref{normal_prior_t}) 
and (\ref{normal_prior_w}) are equivalent up to transposition of the data $\vX$.
To see that consider transposition of model~(\ref{general_model_pca}) $\vX^T -\bmu^T = \vW \vT^T + E^T$. Now the equivalence follows directly from the symmetry of prior distributions for rows of $\vT$ and $\vW$.
Simulation results we present in Section~\ref{sec:simulation-small-big-n} confirm that depending on the relationship between $n$ and $p$, 
one should choose the model dedicated for either $p$ or $n \to \infty$.

\noindent In case of the first prior~(\ref{normal_prior_w}) $PESEL_p$ takes the form:

\begin{align}\label{new_bic_formula}
PESEL^{hetero}_p = \; \log p(\vX | \hat{\mu}, \hat{\vT}, \hat{\sigma}^2) - \log(p) \; &\frac{nk-\frac{k (k+1)}{2} + k + n + 1}{2} \nonumber \\
 = -\frac{pn}{2} \log(2\pi) -\frac{p}{2} \sum_{j=1}^{k} \log \lambda_j  &- 
 \frac{p (n-k)}{2} \log(\hat{\sigma}^2)  -\frac{pn}{2} \\
 & - \log(p) \frac{nk-\frac{k (k+1)}{2} + k + n + 1}{2}, \nonumber
\end{align}

\noindent where 
$\hat{\sigma}^2 = \frac{\sum_{j=k+1}^{n} \lambda_j  }{n-k}$ and $\lambda_j$ 
are singular values in decomposition of covariance matrix for $\vX^T$. 
In case of the second prior distribution~(\ref{normal_variance_prior_w}), formula has the following form:

\begin{align}\label{rajan_bic_formula}
PESEL^{homo}_p = \log p(\vX |\hat{\mu}, \hat{\vU}, \hat{\sigma}^2, \hat{\alpha}) - &\log(p) \frac{nk-\frac{k (k+1)}{2} + n + 1 + 1}{2} \nonumber \\
 = -\frac{p n}{2} \log(2\pi)  -\frac{p k}{2} \log&\left(\frac{\sum_{j=1}^k \lambda_j}{k}\right) 
 -\frac{p (n-k)}{2}  \log(\hat{\sigma}^2) -\frac{pn}{2} \\ 
  & - \log(p) \frac{nk-\frac{k (k+1)}{2} + n + 2}{2}. \nonumber
\end{align}

\section{Simulation study}\label{sec:simulation}

We tested the performance of various model selection methods by comparing distributions of recovered dimensionality,
for data drawn from a known model.
Firstly we aimed to verify how different in practice are
heterogeneous and homogeneous PESEL.
Secondly, how crucial is the assumption of particular asymptotics, i.e.
how much better we can do by using $PESEL_p$ when number of variables exceeds number of observations.
Thirdly, we focused on how robust is PESEL in comparison to state-of-the-art approaches.

\subsection{Methods}\label{sec:simulations_methods}

We present simulation results for seven methods for the estimation of the number of PCs. 
Three of them were previously described in this paper:
\begin{itemize}
 \item \textit{Heterogeneous} PESEL for $n>>p$, $PESEL^{hetero}_n$  defined in formula~(\ref{BIC_model_PPCA}),
and equivalent to BIC for PPCA model proposed by ~\citet{Minka00automaticchoice}.
\item \textit{Heterogeneous} PESEL for $p>>n$, $PESEL^{hetero}_p$ defined in formula~(\ref{new_bic_formula}).
\item \textit{Homogeneous} PESEL for $p>>n$, $PESEL^{homo}_p$ defined in formula~(\ref{rajan_bic_formula}).
\end{itemize}

\noindent We compare those three criteria to four state-of-the-art methods:
\begin{itemize}
 \item Laplace evidence~\citep[eq. 76]{Minka00automaticchoice}, which can be viewed as an extension of $PESEL_n^{hetero}$,
as it contains more terms from Laplace approximation. 
As \citet{Minka00automaticchoice} used a specific non-informative prior distribution on elements of SVD decomposition
of matrix $\vW$ and variance of the noise $\sigma^2$, Laplace evidence depends on that choice and is less general than $PESEL$.
\item Generalized Cross-Validation~\citep{Josse20121869}, which accordingly to the simulation study presented in
\citep{Josse20121869}  performs very well in comparison to many other up-to-date
methods for estimating the number of principal components.
We used implementation from R package FactoMineR~\citep{factominer}.
\item CSV~\citep{2014arXiv1410.8260C}, which is
an exact distribution-based method for testing hypothesis about the number of principal components.
We used our own implementation in MATLAB since the authors did not provide the code for CSV. 
In the simulation study we experienced numerical difficulties with
computing multidimensional integrals that are part of the test statistic. This was observed with a moderate increase 
in either number of variables or signal to noise ratio (defined thereafter). 
CSV assumes that the variance of the noise $\sigma^2$ is known and then it
provides exact testing for the number of principal components. In case when $\sigma^2$ needs to be estimated CSV does not longer guarantee the control of the type I error. To compare CSV with other methods which do not require the knowledge of $\sigma$, we followed the
suggestion made by~\citet{2014arXiv1410.8260C} and estimated $\sigma^2$ by cross-validation
using softImpute R package~\citep{softImpute}.
\item Method proposed in~\citep{passemierestimation}, which uses random matrix theory for 
estimating variance of the noise. This enhanced estimator is then applied for choosing the number of principal components
using Stein's unbiased risk estimator (SURE) or determination criterion of~\citet{bai2002determining}. 
The method is developed under the asymptotic setting when both $n$ and $p$ go to  infinity, and $n/p\rightarrow \gamma>0$.
Implementation is available on author's webpage. 
In simulation results we shall refer to it as Passemier. In our simulations  we use the version of Passemier based on the determination criterion since the software for SURE requires $n>p$, .
\end{itemize}

\noindent Apart from both versions of $PESEL_p$ all other methods are based on the decomposition of the standard covariance matrix, which implicitly assumes the model with independent rows and centers the data by subtracting the columns' means.

\subsection{Simulation scenarios}\label{sec:simulation_scenarios}

\noindent In the simulations, we compared performance for different number of variables in data set, varying from 50 to 2000,
number of observations equal to 50, 100 or 2000,
and signal to noise ratios (SNR) from range [0.25; 8]. 
By \textbf{SNR} we mean the ratio between $l_2$ norm of the columns of the signal matrix $\vM$ and the variance of the noise.
In simulations, we standardized columns of signal matrix $\vM$ to have zero mean and a unit $l_2$ norm, and so
SNR is given by:
$$
\text{SNR} = \cfrac{1}{\sigma^2},
$$
where $\sigma^2$ is the variance of the noise~(as in~(\ref{pca_general_model_hoff})).
Naturally, when the number of variables grows, the signal combined from all variables is relatively higher.
Therefore,  we expect that the performance of different statistical methods should become more accurate when $p$ increases.
This intuition is backed up by simulation results.
\vspace{0.2cm}

We studied four scenarios:

\begin{list}{Scenario \arabic{qcounter}.~\\}{\usecounter{qcounter}}

\item In the first scenario we verified how different are criteria $PESEL^{hetero}_n$~(\ref{BIC_model_PPCA}) and $PESEL^{homo}_n$~(\ref{BIC_model_Rayner}) in practice. In the first scheme we set all non-zero singular values equal to each other:

\begin{algorithm}[H]
\caption{Simulation scheme for signal matrix with equal singular values}
\label{alg:equal_sing_vals}
\begin{algorithmic}[1]
\REQUIRE Number of observations $n$, number of variables $p$, number of PCs $k$, SNR
\STATE Every entry of matrix $\vM$ is drawn from standard normal distribution, $m_{i j} \sim N(0,1)$. 
\WHILE {all singular values in normalized matrix $\vM$ are not equal each other}
\STATE Perform SVD decomposition of matrix $\vM=\vU \vL \vV^T$.
   \STATE Set all first $k$ singular values from $\vL$ equal to their mean and the rest of singular values to 0. 
\begin{align*}
 \tilde{l_i} &:= \frac1k \sum_{j=1}^k l_j, \;\; i=1,\dots, k,\\
\tilde{\vU} &:= \vU[\bigcdot, 1:k] \\
\tilde{\vV} &:= \vV[\bigcdot, 1:k], 
\end{align*}
where $l_j$ is $j$-th element on diagonal of $\vL$.
  \STATE Set $\vM := \tilde{\vU} \tilde{\vL} \tilde{\vV}^T$.
  \STATE Standardize $\vM$ so each column has a zero mean and a unit $l_2$ norm.  
\ENDWHILE 
\STATE $x_{i, j} := m_{i, j} + \N(0, \frac{1}{SNR})$
\end{algorithmic}
\end{algorithm}

The reason for the {\bf while} loop is that after standardization eigenvalues might no longer be equal. Therefore we need several steps to obtain the matrix $\vM$ which has all eigenvalues equal and at the same time it has standardized columns.

\item Second scheme is analogous, but this time we make non-zero singular values decrease exponentially.

\begin{algorithm}[H]
\caption{Simulation scheme for signal matrix with exponentially decreasing singular values}
\label{alg:exp_sing_vals}
\begin{algorithmic}[1]
\REQUIRE Number of observations $n$, number of variables $p$, number of PCs $k$, SNR
\STATE Every entry of matrix $\vM$ is drawn from standard normal distribution, $m_{i j} \sim N(0,1)$ 
\STATE Perform SVD decomposition of matrix $\vM=\vU \vL \vV^T$.
   \STATE Set all singular values of order greater than $k$ to 0 and the largest $k$ to: 
\begin{align*}
\tilde{l_i} &:= C2^{-i}, \;\; i=1,\dots, k \\
\tilde{\vU} &:= \vU[\bigcdot, 1:k], \\
\tilde{\vV} &:= \vV[\bigcdot, 1:k], 
\end{align*}
where $l_j$ is $j$-th element on diagonal of $\vL$ and
$C = (\sum_{j=1}^k l_j)/(\sum_{i=1}^k  2^{-i})$ is a normalizing constant.
  \STATE Set $\vM := \tilde{U} \tilde{L} \tilde{V}^T$
  \STATE Standardize $\vM$ so each column has a zero mean and a unit $l_2$ norm.  
\STATE $x_{i, j} := m_{i, j} + \N(0, \frac{1}{SNR})$
\end{algorithmic}
\end{algorithm}

 \item Data is generated according to the 
fixed effect probabilistic model~(\ref{general_model_pca}). Both scores $\vT$ and 
coefficients $\vW$ are drawn once from multivariate normal distribution:
$\vt_{i\bigcdot } \sim \N(0, \I)$, $\vw_{j\bigcdot } \sim \N(0, \I)$. 
Signal matrix is calculated as $\tilde{\vM} := \vT \vW^T$ and standardized so each column has a zero mean and a unit $l_2$ norm.
In each iteration of the experiment a random noise
is added to the signal matrix ${\vM}$:
\begin{align}
 x_{i,j} &=  {m}_{i,j} +  \epsilon_{i j}  \;\; i = 1,\dots, n \; , \;  j = 1,\dots, p \\
 \epsilon_{i j} &\sim \N(0, \frac{1}{SNR}). \nonumber
\end{align}
\item Data is generated as in Scenario 2. However,
noise is drawn from the rescaled Student distribution with three degrees of freedom $\epsilon_{i j} \sim \frac{1}{SNR} \sqrt{\frac13} t(3)$. 

\item Data is generated as in Scenario 2. However, a number of surplus noisy variables 
$z_i \sim \N(0, \I),\; i = 1, \dots, p/2$ is added to the data. 
$\vX_{n \times \frac32 p}= \left[ \vM_{n \times p} + E_{n \times p} \text{ \textbrokenbar \,} \vZ_{n \times p/2} \right]$.
An example when such violation of our assumptions could occur, is when PCA is used 
in iterative clustering procedure. It might happen that some elements could be falsely classified, 
yet we would still like to retain true dimensionality.

\end{list}

\noindent We replicated each simulation scenario 100 times to get a reliable comparison
between the methods. 

\subsection{Results}

\noindent In the following sections, we present only some selected, yet representative, simulation results. 
True number of principal components is 5. Results for the number of components equal to 2 or 10 were similar, and
therefore are not reported in this paper. We also simulated data from random effects model. Factors and coefficients were drawn
from normal, heavy-tailed (student), skewed (exponential) or uniform distributions. The qualitative conclusions were also
consistent with simulations results presented in this paper.

\subsubsection{Criteria comparison}\label{sec:criteria_comparsion}

\begin{figure}[htp] 
\centering
  \includegraphics[scale=0.43]{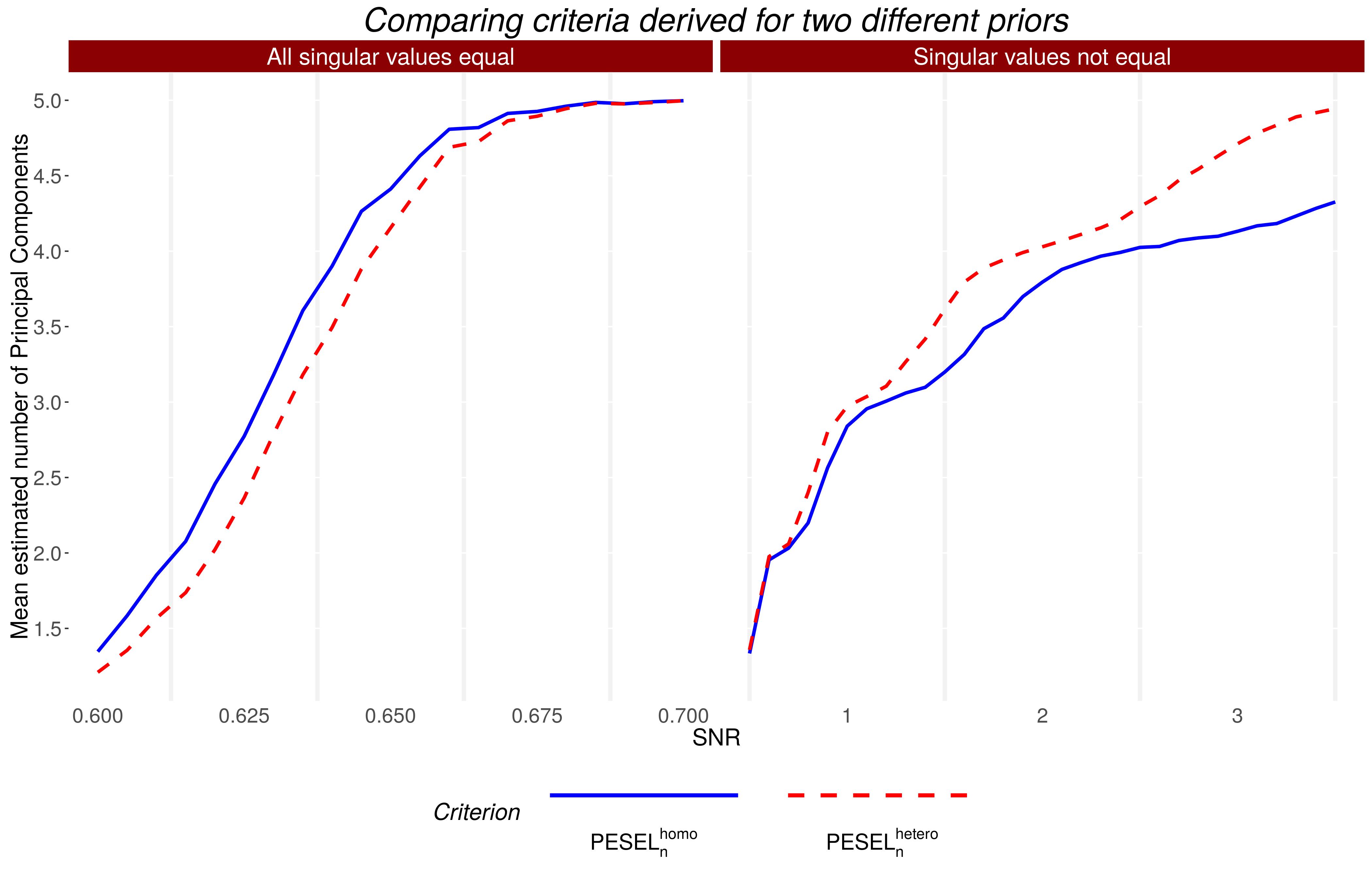}
      \caption{\small Comparison of performance for $PESEL^{hetero}_n$~(\ref{BIC_model_PPCA}) and $PESEL^{homo}_n$~(\ref{BIC_model_Rayner}). 
      Number of variables is 50, number of observations is 100. True number of PCs is 5.
      When singular values are equal, then \textit{homogenous} PESEL has an edge over \textit{heterogenous} PESEL and vice versa.}
      \label{fig:equalNonequalSingularVals}
\end{figure}

As it can be seen in Figure~\ref{fig:equalNonequalSingularVals} the difference in performance between two PESEL criteria 
backs up remarks made in Section~\ref{sec:ppca_bic}. 
$PESEL^{homo}$, which assumes equality of singular values, performs consistently better
when we simulate our data in accordance with this assumption. Contrary, when singular values are not equal, then
$PESEL^{hetero}_n$  gets an edge and
the difference between the methods is larger. Because those two criteria perform comparable, we
shall from now on report only results for $PESEL^{hetero}$.

\subsubsection{Data drawn according to the Scenario 2} \label{sec:simulation-small-big-n}
 
\begin{figure}[htp] 
  \begin{center}
  \includegraphics[scale=0.45]{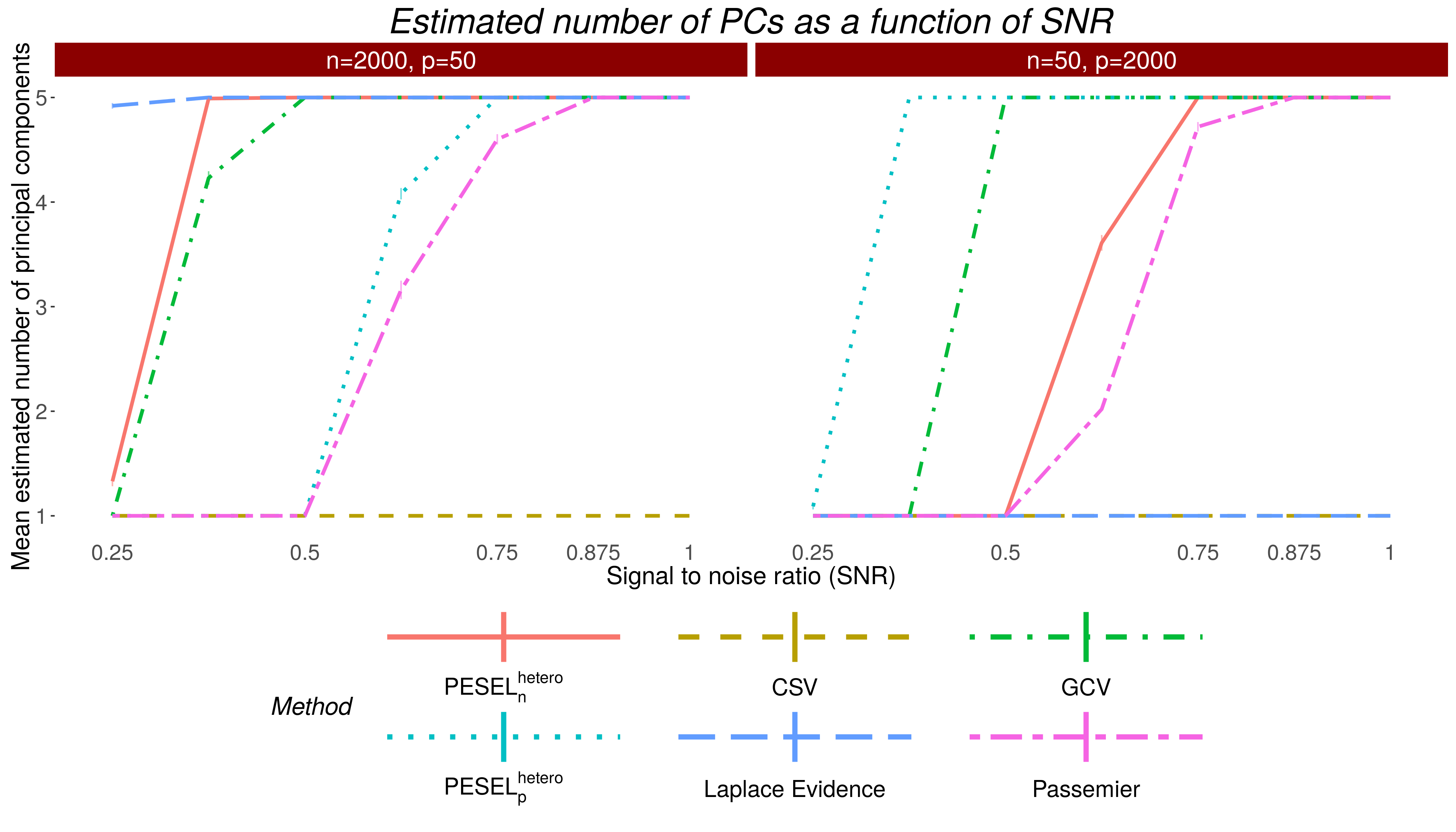}
\caption{\small Data generated according to the Scenario 2. 
True number of components is 5. Results are for $n=2000$, $p=50$ and $n=50$, $p=2000$.}
      \label{fig:big_small_n}
    \end{center}
\end{figure}

Figure~\ref{fig:big_small_n} illustrates the performance of different methods when number of observations $n$ is either very large or very small compared
 to the number of variables $p$.
Results are not surprising, as methods that assume asymptotics in $n$ 
work better when $n$ is large and vice versa. Note, that probabilistic methods outperform GCV when $\frac{p}{n}$ ratio
is in accordance with their underlying asymptotics. In particular $PESEL_p$ is superior to all other approaches when $p>>n$.
In case when $n>>p$ we observe a superior performance of the criterion of \citep{Minka00automaticchoice} based on extended version of Laplace approximation.
 
\begin{figure}[htp] 
   \begin{center}
  \includegraphics[scale=0.45]{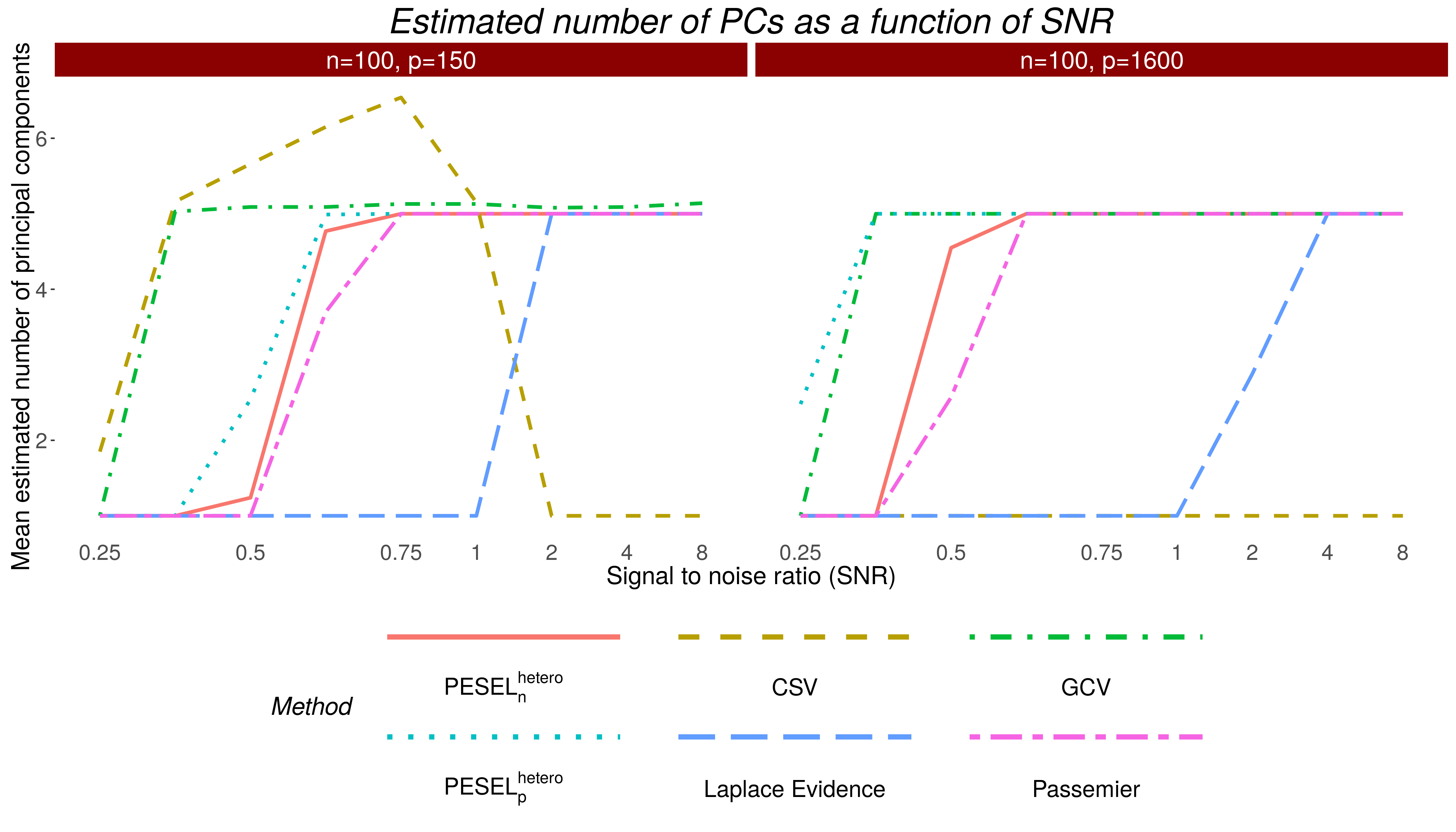}
\caption{\small Data generated according to the Scenario 2.
True number of components is 5.
Results are for number of variables 150 and 1600.
Number of observations is constant and equal 100.}
      \label{fig:model_fixed_snr}
          \end{center}
\end{figure}

\vspace{0.5cm}
\noindent Figure~\ref{fig:model_fixed_snr} illustrates the situation when data is drawn according to the Scenario 2, but with number of variables and observations
more balanced.
As expected $PESEL^{hetero}_n$ is slightly worse than $PESEL^{hetero}_p$ when $p>n$. 
GCV works better when signals are weak and number of variables is comparable to 
number of observations.
Passemier's discriminant criterion is inferior to both PESELs.
Laplace evidence performs poorly when the number of variables is big compared to the number of observations.
CSV works well with weak signals and small number of variables, however, when either one of those grows, it encounters numerical problems 
described in Section~\ref{sec:simulations_methods}.

\subsubsection{Robustness}

As mentioned in the Introduction, the main motivation for testing robustness is when PCA is used as an auxiliary technique. 
In such a case it might have to deal with data with an excessive noise. We report results for two kinds of violations of assumed probabilistic model, previously described in Section~\ref{sec:simulation_scenarios}.

 \begin{figure}[htp] 
  \begin{center}
  \includegraphics[scale=0.40]{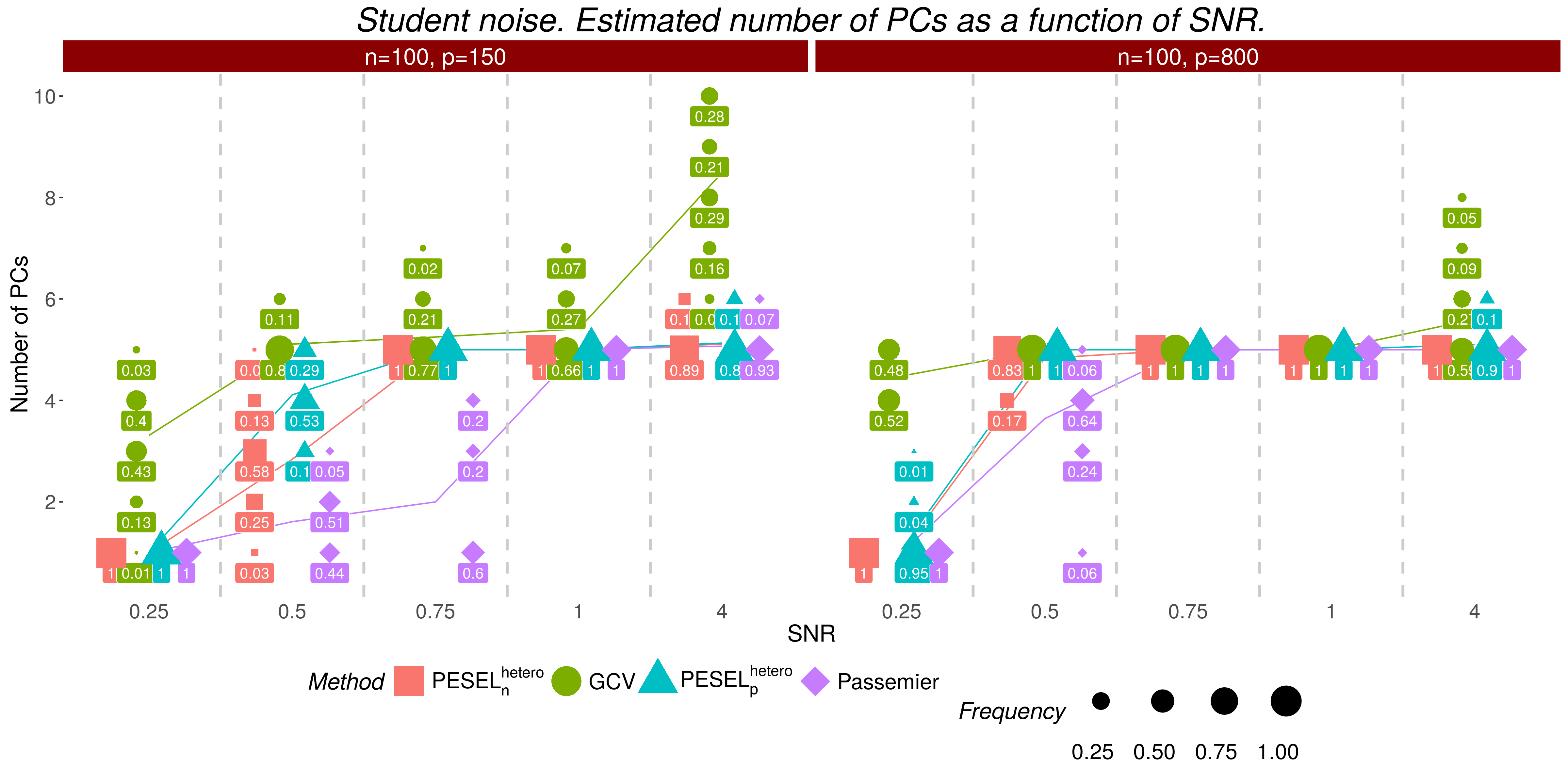}
\caption{\small Data drawn according to the Scenario 3 (noise from Student distribution).
True number of components is 5.
Size of the symbol is proportional to the particular results frequency.
Lines represent mean estimated number of Principal Components.
Results are for number of variables 150 and 800.
Number of observations is constant and equal 100.
}
    \label{fig:student_vars_full}
    \end{center}
\end{figure}

\begin{figure}[htp] 
  \begin{center}
  \includegraphics[scale=0.40]{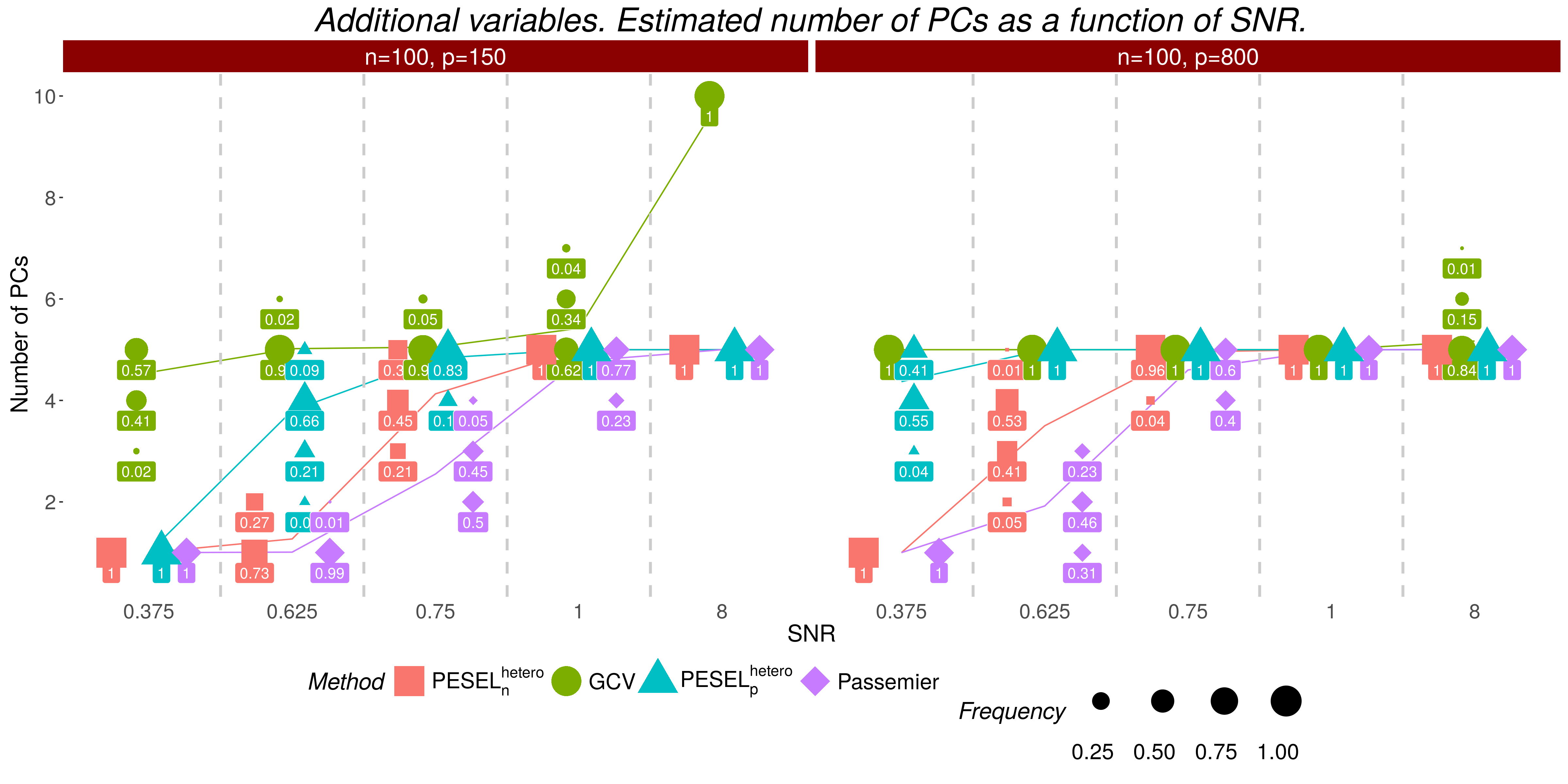}
\caption{\small Data drawn according to the Scenario 4 (surplus noisy variables).
True number of components is 5.
Size of the symbol is proportional to the particular results frequency.
Lines represent mean estimated number of Principal Components.
Results are for number of variables 150 and 800.
Number of observations is constant and equal 100.
}
    \label{fig:additional_vars_full}
    \end{center}
\end{figure}

\begin{figure}[htp] 
  \begin{center}
  \includegraphics[scale=0.40]{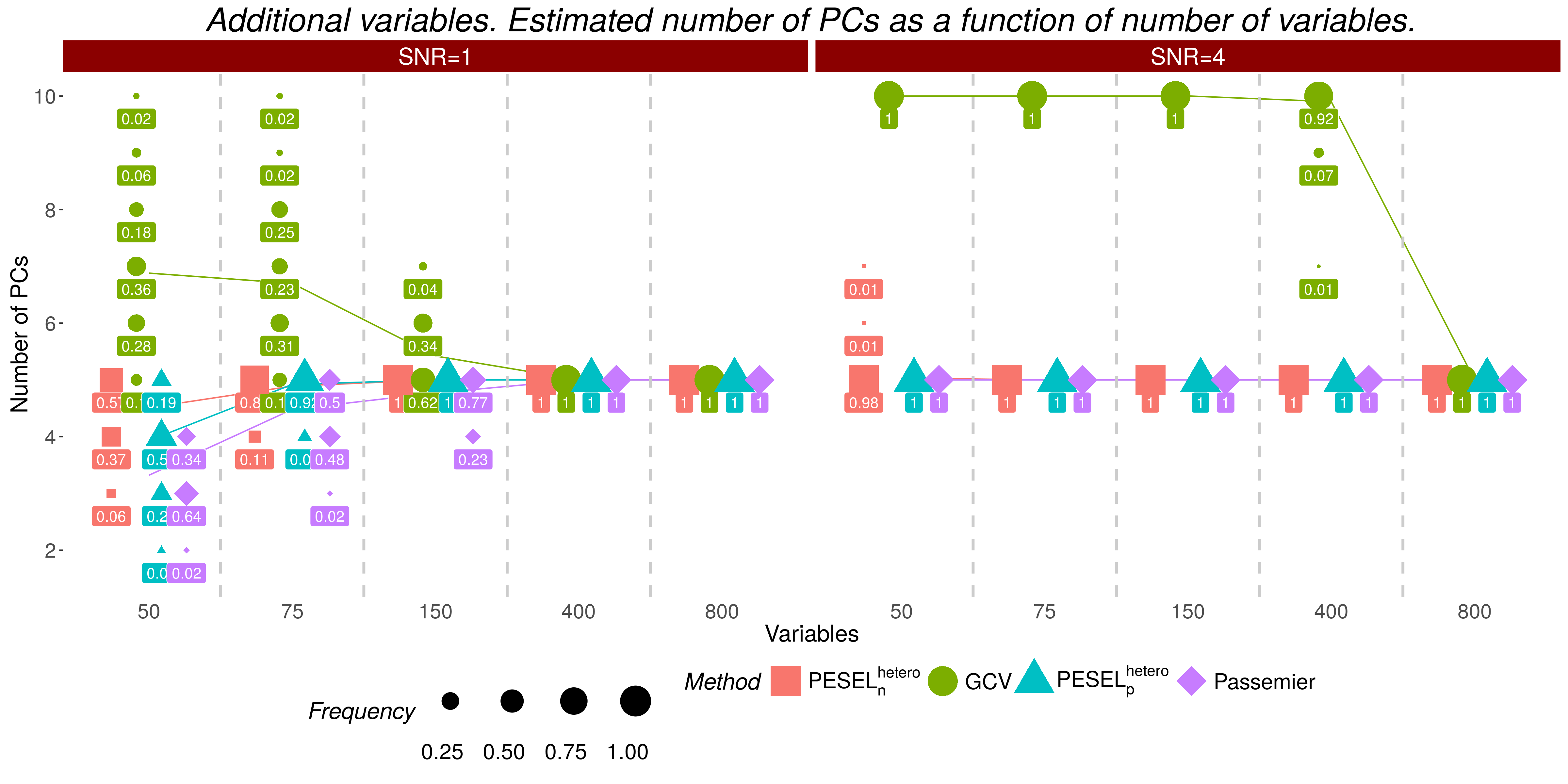}
  \caption{\small Data drawn according to the Scenario 4 (surplus noisy variables).
True number of components is 5.
Size of the symbol is proportional to the particular results frequency.
Lines represent mean estimated number of Principal Components.
Results are for signal to ratios 1 and 4. Number of variables varies from 50 to 800.
Number of observations is constant and equal 100.
}
    \label{fig:additional_snr_full}
    \end{center}
\end{figure}

\noindent For the clarity of plots, in Figures~\ref{fig:student_vars_full},~\ref{fig:additional_vars_full} and~\ref{fig:additional_snr_full} 
we selected only 4 methods for a detailed comparison. 
We observe that in case of violations of model assumptions GCV has a tendency to overestimate the number of components when signal
gets stronger. Passemier is consistently inferior to $PESEL$.
$PESEL_{p}^{hetero}$ is inferior to GCV when signal is weak, but does not overestimate 
the number of principal components when either number of variables or SNR grows.

\noindent As for the methods not included in the plots, $PESEL_{p}^{homo}$ behaves comparably to $PESEL_{p}^{hetero}$.
Laplace evidence proves to be least robust, as it has a tendency to underestimate number of PCs when
probabilistic model is violated. It is also highly dependent on assumed asymptotics i.e. $n>>p$. 
For CSV, when signal gets strong or number of variables gets large, it is becoming increasingly difficult to compute any of
the integrals that this method is based on. As a result, we did not manage to use this method to estimate the number of PCs under such scenarios.

\subsubsection{Simulations results summary}

\noindent All in all, $PESEL_p$ performance is competitive to up-to-date methods. 
Note that GCV is a serious competitor for data drawn according to Scenario 2,
however, when $p$ is large compared to $n$, which is our main focus, $PESEL_p$ is better. 
Similar conclusions come from robustness study, 
despite the fact that $PESEL_p$ was derived under specific probabilistic assumptions.
When number of variables is large compared to
number of observations, its performance is superior to competing methods.
With small number of variables and strong signal it is less prone to overfitting than GCV. 

\subsection{Real data example} \label{sec:real_data}

We use dataset \textit{UrineSpectra} from~\citep{MetabolAnalyze} $\R$ package, 
which contains NMR metabolomic spectra from urine samples of mice.
This datasets consist of 18 observations of 189 variables.
Mice are from two groups, treatment and control group. 
We compared BIC for PPCA programmed in this package, equivalent to $PESEL^{hetero}_n$,
with $PESEL^{hetero}_p$.

\begin{figure}[htp] 
     \begin{center}
  \includegraphics[scale=0.5]{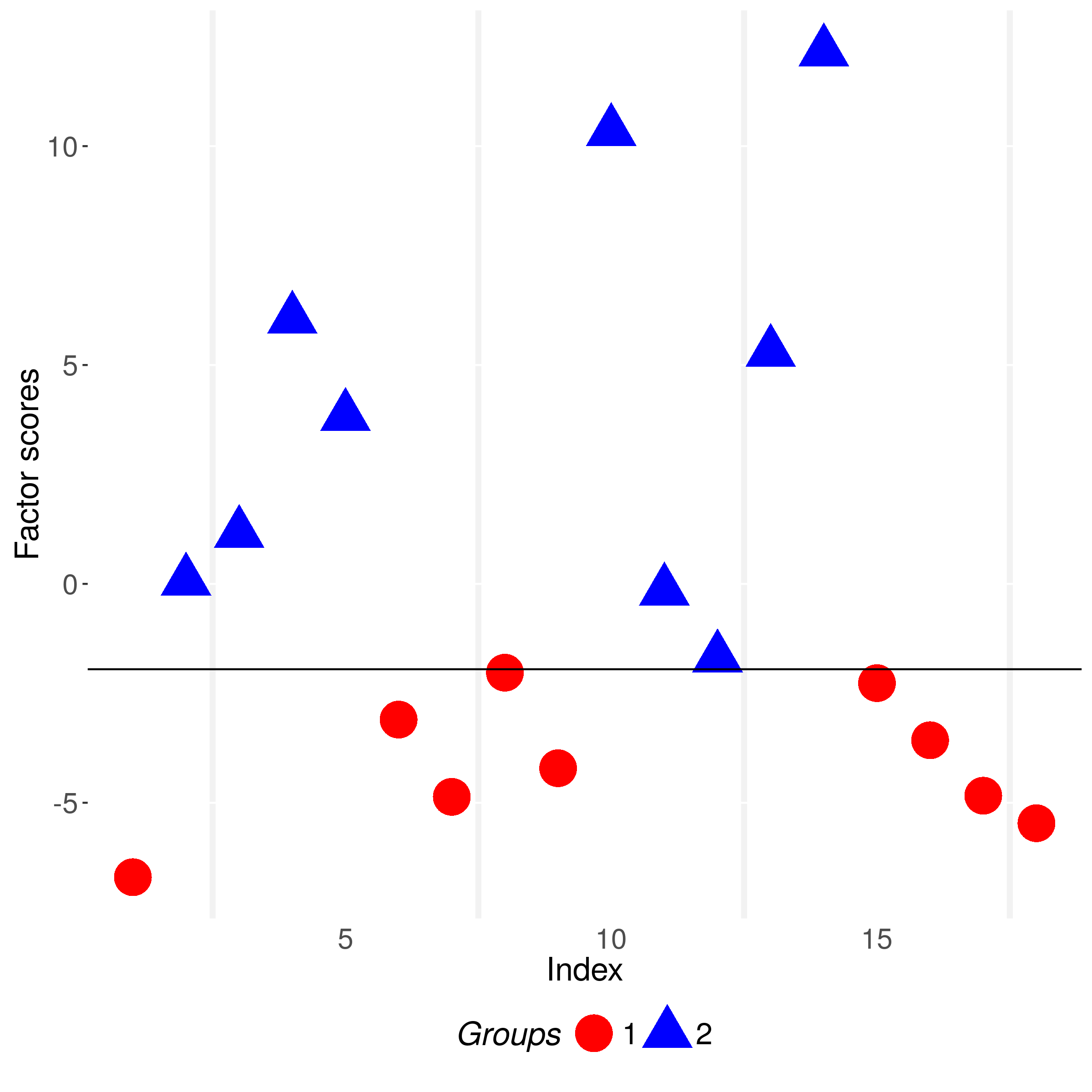}
      \caption{PC scores versus treatment group}
      \label{fig:urineData}
        \end{center}
\end{figure}

\noindent Observe in Figure~\ref{fig:urineData} that first principal component scores allow for perfect discrimination between 
two groups (treatment and control). 
In fact, our method chooses one principal component, while BIC for PPCA
suggests that the true number of principal components is two.
It means that our criterion is able to choose smaller and yet sufficient number of PCs.
In other words, it provides parsimonious model by more restrictive, at least in this case, dimensionality reduction.
In this data set the number of variables is substantially larger than the number of observations ($p>>n$), therefore 
this results is in accordance with intuitions and simulations presented previously.

\section{Conclusion}

\noindent  We presented Bayesian approach for selecting the number of principal components in PCA.
Starting from a fixed effect model~(\ref{general_model_pca}), we presented a framework for 
approximating posterior probability.
As number of parameters in model~(\ref{general_model_pca}) is too big to apply Laplace approximation, 
we suggest imposing prior on either matrix $\vT$ of PCA scores or matrix
$\vW$ of PCA loadings. Obtained PEnalized SEmi-integrated Likelihood ($PESEL$) is valid when either
number of observations or number of variables tends to infinity, while the latter remains fixed. 
We compared performance of derived criteria with state-of-the-art methods in simulations. 
Although $PESEL$ assumes specific probabilistic model,
simulation results suggest that it is more robust than existing methods, especially
in a setting $p>>n$.

\noindent $PESEL$ can be used as an automatic tool for selecting the number of PCs.
One of possible applications are clustering algorithms similar to~\citep{ksubspaces}.
$PESEL$ is implemented in the free software R~\citep{Rlang} package \textit{varclust}~\citep{manualvarclust}, 
where it is used in an iterative variables clustering procedure.

\noindent $PESEL$ works in a specific asymptotics when one of $p$ and $n$ is fixed while latter tends to infinity. 
In case when both these parameters diverge it is not longer justified to use the Laplace approximation. In case when $p$ and $n$ diverge at the same rate \citet{hoyle2008automatic} proposed an approximate Bayesian solution which uses more terms in the asymptotic expansion of the integrated likelihood. Contrary to PESEL this method is dependent on the selection of priors on both $T$ and $W$. We did not consider this approach in our simulation study because of the lack of the publicly available implementation. However, it is worth mentioning that the method of ~\citep{passemierestimation} is also designed to work under the assumption that $p$ and $n$ diverge to infinity at the same rate and that PESEL proved to be consistently superior than~\citep{passemierestimation}
in a range of dimensions that was of our interest.


\noindent Finally, we would like to note that our approach can be easily extended for the cases when matrix $\vX$ is binary or its elements can take on only integer values. Such data can be modeled using the framework of generalized linear models and the respective Bayesian approach
for PCA was proposed e.g. in \cite{Hoff07modelaveraging}. Verification of the performance of PESEL under this setting remains an interesting topic for a further research.

\vspace{0.5cm}

\textbf{Acknowledgment}

\noindent We thank professor Jean-Michel Marin from University of Montpellier for assistance and for comments that improved the manuscript.
PS and MB were
supported  by  the  European  Union’s  7th  Framework
Programme  for  research,  technological  development
and demonstration under Grant Agreement no 602552, cofinanced by the Polish Ministry of Science and Higher Education under Grant Agreement 2932/7.PR/2013/2.
Calculations have been carried out in Wroclaw Centre for Networking and Supercomputing (\url{http://www.wcss.pl}), grant No. 347.

 \clearpage
\bibliographystyle{kbib}
{
\bibliography{varclust_paper}
}

\newpage

\appendix

\section{Laplace approximation} \label{app:laplace}
Laplace approximation is a technique for rough calculation of special kind of integrals that are 
difficult or impossible to compute analytically.
We shall present its multivariate form for the logarithm of the integral

\begin{equation}\label{general_laplace_approximation_formula}
 \log \left(\int_{\Rs^d} g(\theta) e^{-t h(\theta)} d\theta \right) \rightarrow
 t h(\hat{\theta}) + \frac{d}{2} \log{ 2\pi} + \frac12 \det \Sigma_{\hat{\theta}} - \frac{d}{2} \log{t},\;\;\mbox{as}\;\;t\rightarrow\infty\;\;,
\end{equation}
where $\hat{\theta} = \argmax_{\theta} \{-h(\theta) \}$ and $ \Sigma_{\hat{\theta}}$ is a Hessian of $h(\bigcdot)$ evaluated at $\hat{\theta}$. The above holds  
under some mild regularity conditions for $g(\cdot)$ and $h(\cdot)$ described in detail in e.g.~\citep{de1970asymptotic}.

\noindent If we omit terms that do not depend on $t$ we get the approximation
\begin{equation}\label{laplace_approximation_formula}
 \log \left(\int_{\Rs^d} g(\theta) e^{-t h(\theta)} d\theta \right) \approx t h(\hat{\theta}) - \frac{d}{2} \log{t}.
\end{equation}

\noindent In the context of statistical model selection let $ \vY = \{ Y_1, ..., Y_N\}$ be iid sample drawn from 
the distribution with the density or mass function $P(y |M, \theta)$, where $M$ denotes a given model and $\theta$ -- parameters in that model. Moreover, let $\pi_M(\theta)$ be the prior distribution on those parameters. 
In case when all considered models have the same prior probability, the model selection according to the Maximum A Posteriori rule (MAP) is based on maximization of the marginal likelihood of the data $P(\vY|M)$, given by the formula
  \begin{align}
   \log{P(\vY|M)} &= \log \left( \int_{\Theta} P(\vY|M,\theta) \, \pi_M(\theta) \, d\theta \right) \nonumber \\
    &= \log \left( \int_{\Theta} \pi_M(\theta) \exp \left\{ -N  \left( -\frac{1}{N} \log P(\vY|M,\theta)\right) \right\} d\theta \right) \nonumber \\
    &= \log \left( \int_{\Theta} \pi_M(\theta) \exp \left\{ -N  \left( -\frac{\sum_i \log P(Y_i|M,\theta)}{N} \right) \right\} d\theta \right). \nonumber
  \end{align}
To use Laplace approximation observe that we can refer directly
to formula~(\ref{laplace_approximation_formula}), where
$N$ corresponds to $t$, the prior $\pi_M(\cdot)$  to $g(\cdot)$ and averaged likelihood $\frac{\sum_i \log P(Y_i|\theta)}{N}$ to $h(\theta)$. 
By performing this approximation we get the classical BIC criterion
\begin{equation}
  \log \left( \int_{\Theta} P(Y|M,\theta) \, \pi_M(\theta) \, d\theta \right) \approx \sum_i \log P(Y_i|M,\hat{\theta}) - \frac{d}{2} \log{N}
  = \log P(\vY|M,\hat{\theta}) - \frac{d}{2} \log{N}. \nonumber
\end{equation}

\section{Number of parameters in semi-integrated likelihood} \label{app:number-parameters}
Let us first decompose matrix $\vW$, as in~(\ref{ml_estimates_PPCA}):

\begin{align}
  \vW &= \vU (\vL - \sigma^2 \I_k)^{1/2} \vR, \nonumber \\
  \vU^T \vU &= \I_k, \nonumber \\
  \vR^T \vR &= \I_k. \nonumber
\end{align}

\noindent This implies the following equality:

\begin{align}\label{r-orthogonality}
 \vW \vW^T + \sigma^2\vI_p &= \vU (\Lambda_k - \sigma^2 \vI_k)^{1/2} \vR (\vU (\Lambda_k - \sigma^2 \vI_k)^{1/2} \vR)^T + \sigma^2 \vI_p = \nonumber \\
 & = \vU (\Lambda_k - \sigma^2 \vI_k)^{1/2} \vR \vR^T (\Lambda_k - \sigma^2 (\vI_k)^{1/2}))^T \vU^T + \sigma^2\vI_p =  \\ 
  & = \vU (\Lambda_k - \sigma^2 \vI_k) \vU^T + \sigma^2\vI_p. \nonumber
\end{align}
\noindent In the above we use the fact, that $\vR$ is orthogonal square matrix.

\noindent Using \ref{r-orthogonality} we can write likelihood~(\ref{posterior_likelihood_ppca}) for whole data as:
\begin{align}\label{likelihood}
 p(\vX | \mu, \vW, \sigma^2) 
 &= \Pi^n_{i=1} p(\vx_{i, \cdot} |\mu, \vW,  \sigma^2) \nonumber \\
 & = (2\pi)^{-pn/2} |\vW \vW^T + \sigma^2 \vI_p|^{-n/2} \exp \left[ -\frac{1}{2} tr((\vW \vW^T + \sigma^2 \vI_p)^{-1} \vS)\right] \nonumber \\
 & = (2\pi)^{-pn/2} |\vU (\Lambda_k - \sigma^2 \vI_k) \vU^T + \sigma^2\vI_p |^{-n/2} \exp \left[ -\frac{1}{2} tr \left( (\vU (\Lambda_k - \sigma^2 \vI_k) \vU^T + \sigma^2\vI_n )^{-1} \vS \right) \right].
\end{align}

We assume that priors on all parameters are independent. Then, since $\vR$ is not part of likelihood~\ref{likelihood}, 
it can be integrated out. So the integral in~(\ref{posterior_likelihood_ppca}) reduces to
\begin{equation}
 \int p(\vX | \mu, \vW, \sigma^2) d\mu \; d\vW \; d\sigma^2 = 
 \int p(\vX | \mu, \vU, \vL, \vR, \sigma^2) d\mu \; d\vU \; d\vL \; d\vR \; d\sigma^2 =
 \int p(\vX | \mu, \vU, \vL, \sigma^2) d\mu \; d\vU \; d\vL \; d\sigma^2.
\end{equation}

\noindent For $\vU$ we have $pk-\frac{k (k+1)}{2}$ parameters -- this is a dimension of $p\times k$ Steifel manifold~\citep{james1954}. 
$\vL$  has $k$ parameters. $\mu$ has $p$ parameters and $\sigma$ is one parameter.

\end{document}